\documentclass[journal,twocolumn,twoside,times,10pt]{IEEEtran}
\usepackage{epsfig,amsfonts,subfigure,color}
\usepackage{graphicx,cite,amssymb,amsmath,mathrsfs,perso}
\begin{document}
\title{Spatially-Coupled Random Access on Graphs}
\author{Gianluigi Liva, Enrico Paolini,
Michael Lentmaier and Marco Chiani
\thanks{
Gianluigi Liva is with Institute of Communication and Navigation of the Deutsches Zentrum fur Luft-
und Raumfahrt (DLR), 82234 Wessling, Germany (e-mail: Gianluigi.Liva@dlr.de).}
\thanks{
Enrico Paolini and Marco Chiani are with DEIS/WiLAB, University of
Bologna, 47023 Cesena (FC), Italy (e-mail: \{e.paolini, marco.chiani\}@unibo.it).}
\thanks{
Michael Lentmaier is with the Vodafone Chair Mobile Communications
Systems, Dresden University of Technology (TU Dresden), D-01062
Dresden, Germany (e-mail: Michael.Lentmaier@ifn.et.tu-dresden.de).
}}%
\maketitle
\thispagestyle{empty} \pagestyle{empty}

%
\begin{abstract}
In this paper we investigate the effect of spatial coupling applied
to the recently-proposed coded slotted ALOHA (CSA) random access
protocol. Thanks to the bridge between the graphical model
describing the iterative interference cancelation process of CSA
over the random access frame and the erasure recovery process of
low-density parity-check (LDPC) codes over the binary erasure
channel (BEC), we propose an access protocol which is inspired by
the convolutional LDPC code construction. The proposed protocol
exploits the terminations of its graphical model to achieve the
spatial coupling effect, attaining performance close to the
theoretical limits of CSA. As for the convolutional LDPC code case,
large iterative decoding thresholds are obtained by simply
increasing the density of the graph. We show that the threshold
saturation effect takes place by defining a suitable counterpart of
the maximum-a-posteriori decoding threshold of spatially-coupled
LDPC code ensembles. In the asymptotic setting, the proposed scheme
allows sustaining a traffic close to $1\,\mathrm{[packets/slot]}$.
\end{abstract}
%
%
\section{Introduction}\label{sec:intro}

Since the introduction of the ALOHA protocol \cite{Abramson:ALOHA},
several random access (RA) {schemes} have been introduced. Among
them, some feedback-free RA protocols originally proposed in
\cite{Massey85:collision_channel,Hui84:CCwoFB} re-gained attention
in the recent past
\cite{Thomas00:Capacity_Wireless_Collision,Shum09:Shift_Invariant_Protocol}.
In \cite{Massey85:collision_channel}, the capacity of the so-called
collision channel without feedback (CCw/oFB) was analyzed, assuming
slot-aligned but completely asynchronous users' transmissions.
Moreover, a simple approach to achieve error-free transmission (in
noise-free setting) over the CCw/oFB was proposed. In the context of
the CCw/oFB, the capacity is defined as maximum packet transmission
rate per slot, which allows the receiver to recover the packets with
an arbitrarily-small error probability (in noise-free conditions).

 The approach of
\cite{Massey85:collision_channel} consists of assigning different
periodic protocol (access) sequences to the users. Each sequence
defines in which slots each user is allowed to access the shared
channel. Furthermore, the users encode their packets by means of
erasure correcting codes. The user's packet can be recovered
whenever a sufficient number of codeword segments are received
collision free. Hence, by selecting proper protocol sequences, it is
possible to ensure that a sufficient number of segments per user are
recovered, even if the beginning of the different protocol sequences
is unsynchronized. In this way, a symmetric capacity\footnote{The
symmetric capacity is given by the sum-rate capacity under the
hypothesis that all users adopt the same transmission rate.} equal
to $1/e\,\mathrm{[packets/slot]}$ is achieved as $N\rightarrow
\infty${,} where $N$ is the number of users accessing the RA
channel. The same capacity is achieved also in the unslotted case.
Although simple, the approach of \cite{Massey85:collision_channel}
poses some challenges, especially if a large (and varying) number of
users has to be served
\cite{Hui84:CCwoFB,Thomas00:Capacity_Wireless_Collision}.

Recently, RA schemes profiting from successive interference
cancelation (SIC) have been introduced and analyzed
\cite{DeGaudenzi07:CRDSA,Giannakis07:SICTA,Liva11:IRSA,Paolini11:CSA_ICC}.
These schemes share the feature of canceling the interference caused
by collided packets on the slots where they have been transmitted
whenever a clean (uncollided) copy of them is detected. In
\cite{Liva11:IRSA,Paolini11:CSA_ICC} it was shown that the SIC
process can be well modeled by means of a bipartite graph. The
analysis proposed in \cite{Liva11:IRSA,Paolini11:CSA_ICC} resembles
{density evolution} analysis of low-density parity-check (LDPC) and
doubly-generalized LDPC (D-GLDPC) codes over erasure channels
\cite{studio3:GallagerBook,studio3:richardson01capacity,paolini10:random}.
By exploiting design techniques from the LDPC context, a
remarkably-high capacity (e.g. up to $0.8\,\mathrm{[packets/slot]}$)
can be achieved in practical implementations. The schemes considered
in \cite{DeGaudenzi07:CRDSA,Giannakis07:SICTA,Liva11:IRSA} assume a
feedback from the receiver to achieve a zero packet loss rate.

A scheme based on the coded slotted ALOHA (CSA) of
\cite{Paolini11:CSA_ICC} has been analyzed in the context of the
CCw/oFB in \cite{Paolini11:CSA_Globecom}. In
\cite{Paolini11:CSA_Globecom} an upper bound on the maximum load $G$
sustainable at a scheme rate $R$, has been derived as the unique
positive solution to
\begin{equation}\label{eq:AREA_CONDITION_FINAL}
G\, =\, 1- e^{-G/R}
\end{equation}
in $[0,1)$. Still in \cite{Paolini11:CSA_Globecom} it was shown how this bound
can be tightly approached by a
careful selection of the distribution of the codes to be used at
users for encoding their packets.

In this paper, we propose another means for approaching the bound
defined by \eqref{eq:AREA_CONDITION_FINAL}, which is based on
spatial coupling. Spatial coupling effects were initially devised in
the context of density evolution analysis of convolutional LDPC
codes over the binary erasure channel (BEC)
\cite{Felstrom99:ConvLDPC_TIT,Lentmaier2004:Allerton_BEC,SpatialCoupling:kudekar_TIT,
Lentmaier2010:ISIT_TerminatedCapacityGLDPC} and the additive white
Gaussian noise (AWGN) channel \cite{Lentmaier2010:AWGN}.
Subsequently, its application to other settings relying on sparse
graph representations has been investigated (see e.g.
\cite{SpatialCoupling:compresensing,Urbanke2010:ITW_CoupledGM,Truhachev2012:ITA_inaircoupling}).
By imposing some constraints on the CSA access scheme, we show how
the threshold under the iterative (IT) SIC process saturates towards
a suitably-defined equivalent of the maximum-a-posteriori (MAP)
decoding threshold of LDPC ensembles.


\section{Coded Slotted Aloha: Erasure Decoding Model}\label{sec:access_and_CSA}

We recall next the basic model adopted for the description of CSA.
We consider {a} slotted RA scheme where slots are grouped in medium
access control (MAC) frames, all with the same length (in slots).
Each user is frame- and slot-synchronous, and attempts {at most} one
\emph{burst} (i.e., packet) transmission per MAC frame. Each burst
has a time duration $T_{\mathrm{slot}}$, whereas the MAC frame is of
time duration $T_{\mathrm{frame}}$. Neglecting guard times, the MAC
frame is composed of $M=T_{\mathrm{frame}}/T_{\mathrm{slot}}$ slots.
We consider a population of $N$ users, with $N\gg M$. {Users} are
characterized by a \emph{sporadic} activity, i.e., at the beginning
of a MAC frame each user {generates} a burst to be transmitted
within the MAC frame {with probability $\epsilon\ll 1$, where}
$\epsilon$ is called \emph{activation probability}. Users attempting
the transmission within {a} MAC frame are referred to as
\emph{active} users. On the contrary, users that are idle during a
MAC frame are referred to as \emph{inactive} users. We denote the
population size normalized to the frame size by $\alpha=N/M$. The
number of active users is modeled by the random variable $N_a$,
which is binomially-distributed with mean value
$\mathbb{E}[N_a]=N\epsilon$. We say that the average offered channel
traffic (representing the average number of bursts transmissions per
slot) is
\[
G=\mathbb{E}[N_a]/M=\epsilon N/M=\epsilon \alpha.
\]

\begin{figure}[]
\begin{center}
\includegraphics[width=0.8\columnwidth,draft=false]{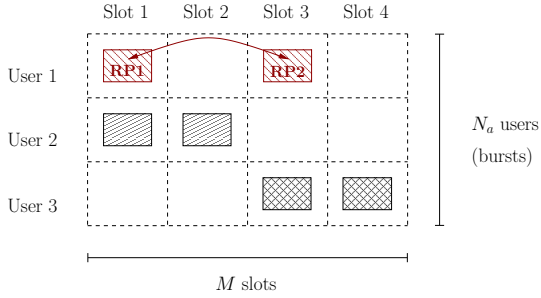}
\end{center}
\caption{MAC frame composed by $M=4$ slots with {$N_a=3$} users attempting a transmission.
Repetition rate $d=2$. }\label{fig:CRA_frame}
\end{figure}

\begin{figure}[]
\begin{center}
\includegraphics[width=0.6\columnwidth,draft=false]{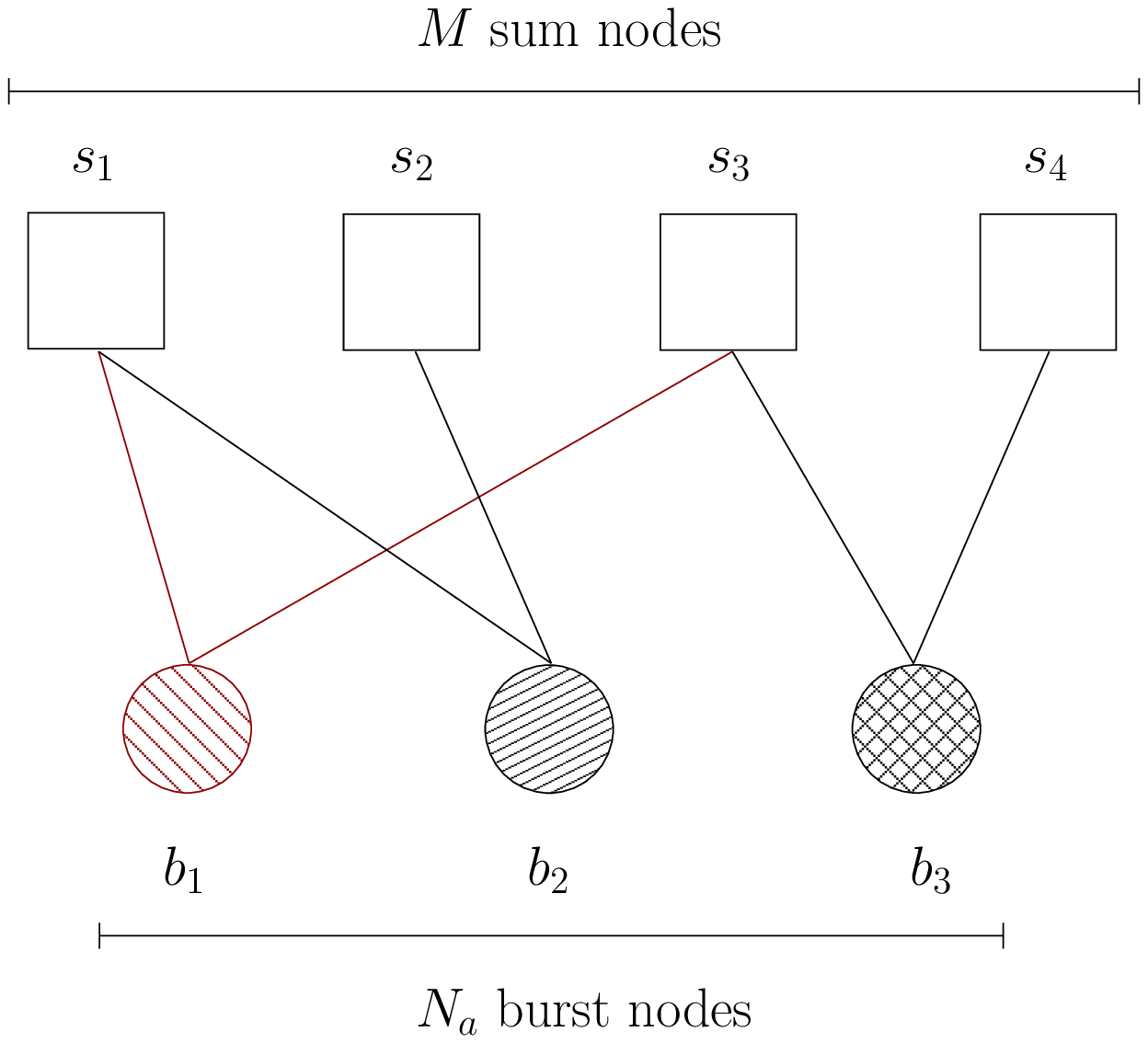}
\end{center}
\caption{{Residual} graph representation for the MAC frame of
Fig.~\ref{fig:CRA_frame}.}\label{fig:CRA_graph}
\end{figure}

We consider a CSA scheme {based on} $(d,1)$ repetition codes, which
is equivalent to a $d$-regular contention resolution diversity
slotted ALOHA (CRDSA) scheme \cite{DeGaudenzi07:CRDSA}. More
specifically, at the beginning of a MAC frame, each user selects $d$
slots with a uniform probability out of the $M$ frame slots. If the
{user} is active, it {transmits} $d$ copies of its burst in the $d$
selected slots. We define $R=1/d$ as the rate of the scheme. In each
burst replica, a pointer to the position of the other copies is
included, e.g., in a dedicated header field. Whenever a
\textit{clean} burst (i.e., a burst which did not collide) is
successfully decoded, the pointer is used to determine the slots
where its copies have been transmitted. Supposing that a another
replica of this burst has collided, it is possible to subtract, from
the signal received in the corresponding slot, the interference
contribution of the twin burst. This may allow the decoding of
another burst transmitted in the same slot. The SIC proceeds
iteratively, i.e., \emph{cleaned} bursts may allow solving other
collisions. {An} example of a MAC frame with $M=4$ slots and $N_a=3$
active users is depicted in Fig.~\ref{fig:CRA_frame}, where the
repetition rate is $d=2$.

Considering a MAC frame composed of $M$ slots and a population of
$N=\alpha M$ users, the frame status can be described by a bipartite
graph, $\msr{G}=(B,S,E)$, consisting of a set $B$ of $N$
\textit{burst nodes} (one for each user), a set $S$ of $M$
\textit{sum} \textit{nodes} (one for each slot in the frame), and a
set $E$ of edges. An edge connects a burst node (BN)
 $b_i\in B$ to a sum node (SN) $s_j\in S$ if and only if the $j$-th slot has been selected by the
$i$-th user at the beginning of the MAC frame. The graph obtained by
expurgating from $\msr{G}$ the BNs associated with inactive users
and their adjacent edges is {called} the \emph{residual graph} and
{is} denoted by $\msr{G}_a=(B_a,S,E_a)$. Here, $B_a\subseteq B$ is
the subset of BNs associated with the active users, and
$E_a\subseteq E$ is the subset of the edges associated with the
transmitted burst copies. An example of the residual graph
representing the MAC frame of Fig.~\ref{fig:CRA_frame} is given in
Fig.~\ref{fig:CRA_graph}.

The SIC process can be represented through a message-passing along
the edges of the graph. As in \cite{DeGaudenzi07:CRDSA,Liva11:IRSA},
we make use of two assumptions which allows simplifying the SIC
process analysis in the graphical model. First, we assume that
perfect SIC is performed. Second, we claim that, whenever a clean
(collision-free) burst is present in a slot, decoding succeeds with
a probability that is essentially $1$. It has been shown in
\cite{DeGaudenzi07:CRDSA,Liva11:IRSA} that these assumptions are
accurate enough to model the SIC process down to low signal-to-noise
ratios (SNRs) with moderate-complexity signal processing algorithms.

Thanks to this simplification, the {SIC} procedure is equivalent to
iterative decoding of {an} LDPC code with $N$ variable nodes and $M$
check nodes over a BEC with erasure probability $\epsilon$
(coinciding with the activation {probability).} All variable nodes
have degree $d$, while the check node degree{s follow} a Poisson
distribution \cite{Liva11:IRSA} with average degree $d N/M=d\alpha$.
The {\emph{nominal code rate}} is {thus} $R_0=1-M/N=1-1/\alpha$.

For large frames ($M\rightarrow\infty$) {and} for a given normalized
population size $\alpha$, CSA shows a threshold behavior. For an
activation probability $\epsilon$ lower than a threshold value
$\epsilon^\mathsf{IT}_\mathsf{block}$\footnote{The subscript
``block'' is here used to emphasize the block-structure of the MAC
frame, in {contrast} with the spatially-coupled structure
{introduced in Section~\ref{sec:convolutional_CSA}}.}, vanishing
burst error probability can be achieved by iterating SIC. The
threshold $\epsilon^\mathsf{IT}_\mathsf{block}$ can be analyzed via
density evolution over the residual graph {$\msr G_a$ according} to
the recursions
\begin{equation}
q_{{\ell}}=p_{{\ell}-1}^{d-1}\label{eq:qi_evol}
\end{equation}
\begin{equation}
p_{{\ell}}=\sum_{{h}} {\tilde{\rho}_{{h}}}
\left(1-\left(1-q_{{\ell}}\right)^{{h}-1}\right)=1-{\tilde{\rho}}\left(1-q_{{\ell
}}\right)\, , \label{eq:pi_evol}
\end{equation}
where ${\tilde{\rho}_h}$ {is} the fraction of edges in $\msr G_a$
connected to SNs with degree ${h}$ in the residual code graph, and
$\tilde{\rho}(x)=\sum_{{h}} \tilde{\rho}_{{h}} x^{{h}-1}$. {In
\eqref{eq:qi_evol} and \eqref{eq:pi_evol}}, $q_{{\ell}}$ and
$p_{{\ell}}$ denote the probabilit{ies} that an edge in the residual
graph carrie{s} an erasure {outgoing from a BN and from a SN,
respectively, at the $\ell$-th iteration}. Since the number of
collisions in a slot follows a Poisson distribution, we have
\begin{equation}
{\tilde{\rho}}(x)=e^{-\epsilon \alpha d(1-x)}.\label{eq:rho}
\end{equation}
Thus, the threshold $\epsilon^\mathsf{IT}_\mathsf{block}$ is
given by the supremum of the set of $\epsilon>0$ such that
\begin{equation}
q>\left(1-e^{-q\epsilon \alpha d}\right)^{d-1} \quad \forall q \in
(0,1]. \label{eq:threshold_cond_ref-eps}\end{equation}

The threshold can be expressed equivalently in terms of offered
traffic. By recalling that $G=\epsilon \alpha$, the threshold
$G^\mathsf{IT}_\mathsf{block}$ is given by the
supremum of the set of $G>0$ such that
\begin{equation}
q>\left(1-e^{-qGd}\right)^{d-1} \quad \forall q \in (0,1]\, ,
\label{eq:threshold_cond_ref}
\end{equation}
{and we have}
$G^\mathsf{IT}_\mathsf{block}=\epsilon^\mathsf{IT}_\mathsf{block}
\alpha$.


\section{Spatially-Coupled CSA: Access Model and Density Evolution}\label{sec:convolutional_CSA}
In this section, we modify the access rules of CSA {to implement} a
convolutional-oriented structure that {enables the exploitation of}
the spatial coupling effect.

\subsection{Access Model}
\begin{figure}[t]
\begin{center}
\includegraphics[width=0.98\columnwidth,draft=false]{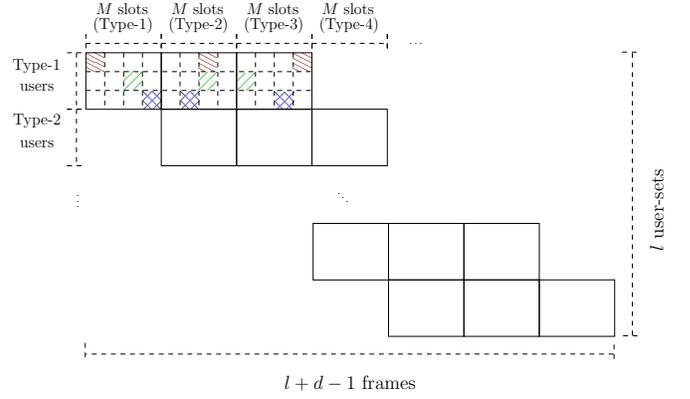}
\end{center}
\caption{Example of a convolutional super-frame structure with $3$ users per user type and $M=4$
slots per frame.}\label{fig:conv_frame}
\end{figure}

The modified access rules are summarized next (see also Fig.~\ref{fig:conv_frame}).
 A super-frame is divided into $M_f=l+d-1$ frames of $M$ slots
each. The slots belonging to the same frame constitute a slot type
set. {A} user {becoming active at the beginning of a frame (with
probability $\epsilon$)} transmits a burst in a slot picked
uniformly at random within {that} frame. Furthermore, {a copy of the
burst is sent} in each of the following $d-1$ frames {in a slot
picked with uniform probability in each frame}. The set of users
{becoming active at the beginning of} the $i$-th frame is referred
to as the {t}ype-$i$ user set. Similarly, the slots belonging to the
{$j$}-th frame are referred to as {type-$j$} slots. The expected
size of a user set is $\mathbb{E}[N_u]=\epsilon N$. Thus, as
before we can define the offered traffic $G$ as
$G=\mathbb{E}[N_u]/M=\epsilon N/M$.

{After} transmission of the $l$-th frame, {transmissions from} new users
{are forbidden}, and the following $d-1$ frames are filled just with the
copies of the bursts whose transmissions have been initiated during
the past $d-1$ frames. Once all the burst copies have been
transmitted, a new transmission cycle begins, i.e.{, a} new super-frame
is initialized.

A (residual) bipartite {graph} description of the recovery process
is obtained as follows. We associate a BN to each {user}. Similarly,
we associate a SN to each slot. The BNs {corresponding} to users of
type $i$ are clustered in {type-$i$ BN groups}, whereas the SNs
related to slots of type $i$ are clustered in {type-$i$ SN groups}.
The number of {BN} types connected to a SN type-{$j$} group is
denoted by $\delta_{{j}}$ (degree of the {type-{$j$} SN group}).
Note that $\delta_{{j}} \in {\{}1,\ldots,d{\}}$. The type-$i$ {BN}
group is said to be neighbor of a type-$j$ {SN} group (and
viceversa) {when} the nodes belonging to the type-$i$ {BN} group are
connected to some nodes in the type-$j$ {SN} group. The indexes of
the groups that are neighbors of the type-$j$ {SN} group {form} the
set $\mathcal{N}^{s}_j$, while the indexes of the groups that are
neighbors of the type-$i$ {BN} group {form} the set
$\mathcal{N}^{b}_i$. Note that the period in which new user
transmission{s} are blocked is equivalent to the termination in the
context of convolutional LDPC codes.\footnote{A loss in terms of
offered traffic, with respect to $G=\mathbb{E}[N_u]/M$, occurs when
the offered traffic is calculated taking into account the frames in
which new arrivals are blocked. Nevertheless, this traffic loss is
negligible for large~$l$.} An example of a super-frame structure is
displayed in Fig.~\ref{fig:conv_frame}. The bursts transmitted into
termination frames experience a lower collision probability than the
other bursts, thus boot-strapping the iterative decoding process
through the coupled structure.

\subsection{Density Evolution}
{Let $p_j$ be the probability that an edge incident on the type-$j$
SN group carries an erasure message towards the BNs, after SN
processing at the generic SIC iteration. Analogously, let $q_j$ be
the probability that an edge incident on the type-$j$ SN group
carries an erasure message towards the type-$j$ SNs, after BN
processing at the generic SIC iteration. Moreover, let
$q_{i\rightarrow j}$ be the probability that an edge emanating from
the type-$i$ BN group carries an erasure message towards the
type-$j$ SN group (with $j\in\mathcal{N}^{b}_i$), after BN
processing at the generic SIC iteration}. The physical load (i.e.,
the load including burst copies) for the $i$-th sub-frame is given
by $G^{(i)}=G\cdot \delta_i${.}

\begin{figure}[t]
\begin{center}
\includegraphics[width=\linewidth,draft=false]{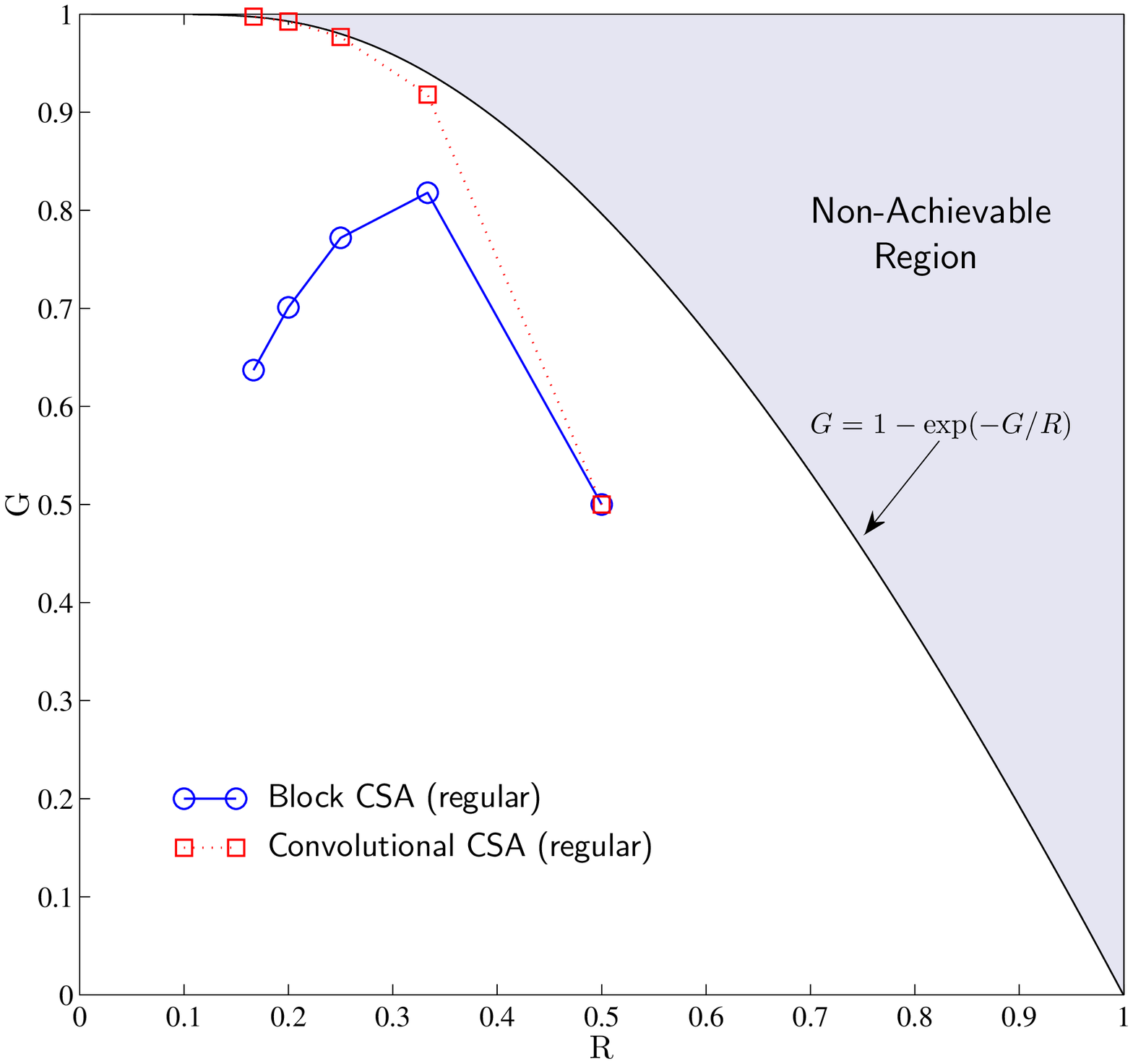}
\end{center}
\caption{Thresholds $G^\mathsf{IT}_\mathsf{block}$ and $G^\mathsf{IT}_\mathsf{conv}$ vs. rate,
$R=1/d$, for CSA based on $(d,1)$ repetition codes.}\label{fig:RG}
\end{figure}

{Next, we define} SN degree distributions {from an edge perspective
as}
\begin{align*}
\rho^{({j})}(x) &= \sum_{{t}=0}^{\infty} \rho^{({j})}_{{t}} x^{{t}-1} \\
\, & =\exp\left(-G\delta_{{j}}(1-x)\right)
\end{align*}
where $\rho^{({j})}_{{t}}$ is the fraction of the edges emanating
from type-${j}$ SNs {and incident on} type-${j}$ SNs with degree
{{$t$}}{.} {D}ensity evolution equations can be {now} derived as
follows{, where $\ell$ is the iteration index}. {For} the {type-$j$
SN group we have}
\[
{p_{j,\ell}}=1-\rho^{({j})}(1-{q_{j,\ell}})
\]
{where}
\[
{q_{j,\ell}}=\frac{1}{\delta_{{j}}} \sum_{v\in\mathcal{N}^S_{{j}}}q_{v\rightarrow
{j,\ell}}\, .
\]
{Moreover, for the type-$i$ BN group, for all $i\in
\mathcal{N}^{b}_j$ we have}
\[
q_{{i\rightarrow j,\ell}}=\prod_{u\in\mathcal{N}^{{b}}_{{i}}\backslash {j}}
{p_{u,\ell-1}}\, .
\]
The SIC IT thresholds for both block-based CSA and its convolutional
counterpart are plotted in Fig.~\ref{fig:RG} versus the bound
\eqref{eq:AREA_CONDITION_FINAL}, as {functions of the rate $R=1/d$}.
{The} thresholds for the {spatially coupled access scheme} are
denoted {by} $G^\mathsf{IT}_\mathsf{conv}$, to {emphasize} the
analogy with convolutional LDPC ensembles. The large SIC IT
thresholds attained by the convolutional CSA scheme allow {to}
tightly {approach}, already for $d=3$, the limit {imposed by}
\eqref{eq:AREA_CONDITION_FINAL}. For $d\geq 4$, an {impressive}
offered traffic, {very close to} $1\,\mathrm{[packet/slot]}$, can be
handled by the convolutional CSA scheme with vanishing packet (i.e.,
burst) loss probability (in the asymptotic setting). The bound for
higher rates $R$ could be tightly approached by spatially-coupled
CSA based on non-trivial $(d,k)$ constituent codes with rate
$k/d>1/2$ \cite{Paolini11:CSA_ICC}.

\section{Threshold Saturation {in CSA}}
We {now introduce} an enhanced decoding algorithm for the
conventional (block) CSA case of Section \ref{sec:access_and_CSA},
which {serves} to derive an upper bound on the achievable threshold
for {CSA schemes, and to investigate threshold saturation effects
for the convolutional scheme}. {This} algorithm mimics the MAP
decoder of an LDPC code over the BEC{, and we refer to it} as
genie-aided maximum-a-posteriori (GA-MAP) decoder.

\subsection{Genie-Aided MAP Decoding}
From an analysis viewpoint, the relation between the transmitted
bursts and the {slot} observations can be simplified by to a matrix
representation of the graph via a{n} $M\times N_a$ binary matrix
$\mathbf{Q}$, where $q_{i,j}=1$ iff {BN} $b_j$ is connected to {SN}
$s_i$ in $\msr G_a$, and $q_{i,j}=0$ otherwise. We denote by
$\mathbf{u}$ the length-{$N_a$} binary vector {whose} $j$-th element
$u_j$ is associated with the modulated burst of user $j$. We {also}
denote by $\mathbf{y}$ the length-$M$ binary vector {whose} $i$-th
element is associated with the $i$-th slot. An equation system
relating $\mathbf{u}$ and $\mathbf{y}$ is thus
\begin{equation}
\mathbf{u}\mathbf{Q}^T=\mathbf{y}.\label{eq:system}
\end{equation}
In this simplified setting, the elements of $\mathbf{u}$ and
$\mathbf{y}$ are binary digits which provide abstraction of the
actual bursts transmitted by the users and the signals received in
the slots, respectively. Upon receiving $\mathbf{y}$ and assuming
that $\mathbf{Q}$ is revealed by a genie, the GA-MAP decoder solves
\eqref{eq:system} for $\mathbf{u}$ via Gauss-Jordan elimination
(GJE). Note that the iterative decoding process described in Section
\ref{sec:access_and_CSA} succeeds only if the matrix $\mathbf{Q}$
can be posed in triangular form by row/column permutations, i.e.{,}
only if the equation system \eqref{eq:system} can be solved
iteratively. Thus, the GA-MAP decoder performance (which is optimum
with respect to \eqref{eq:system}) provides a lower bound on the
decoding error probability of the iterative SIC process.

\subsection{CSA Analysis under GA-MAP Decoding}
We establish next a bridge towards the MAP decoding threshold of
LDPC codes under MAP decoding in order to derive the {threshold} of
a $d$-regular CSA scheme under GA-MAP decoding,
$G^\mathsf{MAP}_\mathsf{block}$. We define $\msr{C}_{d,M,N}$ to be
the ensemble of all length-$N$ codes given by the null space of an
$M \times N$ binary parity-check matrix $\mathbf{H}$, having
constant column weight $d$ and where the {$d$} $1$s in each column
are placed {in random positions, according to} a uniform
distribution{.} Recall that, for the codes in this ensemble, the
nominal rate is given by $R_0=1-M/N$. From a bipartite graph
perspective, the graph of a code in $\msr{C}_{d,{M,N}}$ possesses a
constant variable node degree, $d_v=d$ whereas{, as $N$ and
$M=(1-R_0)N$ tend to infinity,} the check node degree distribution
follows a Poisson distribution with mean value $\overline{d}_c=d
N/M$. {T}he edge-oriented check node degree distribution is {thus}
given by $\rho(x)=\exp(-\overline{d}_c(1-x))$~\cite{Liva11:IRSA}.

Recall that the ensemble under consideration can be placed in
analogy to the scheme introduced in Section \ref{sec:access_and_CSA}
where $N$ is the user population size, $M$ is the number of slots
per frame and $d$ is the repetition rate for the bursts. The IT
decoding threshold $\epsilon^\mathsf{IT}_\mathsf{block}$ over the
BEC for the ensemble $\msr{C}_{d,M,N}$, $N \rightarrow \infty$, is
calculated as the maximum value  of the channel erasure probability
$\epsilon$ (the analogous of the activation probability, in the CSA
context) for which the erasure probabilities $q_i, p_i$ (where $i$
is the iteration index) converge to an arbitrarily-low positive
value, for $i\rightarrow \infty$, according to
\begin{equation}
p_i=\epsilon q_{i-1}^{d-1},
\end{equation}
\begin{equation}
q_i=1-\rho(1-p_i)=1-\exp(-\overline{d}_c p_i).
\end{equation}
The average extrinsic erasure probability $p_e(\epsilon)$  under IT
decoding is obtained finally as
\begin{equation}
p^{\mathsf{IT}}_e(\epsilon)=\lim_{i\rightarrow \infty} q_i^d.
\end{equation}
 Defining an average extrinsic erasure probability function $p^\mathsf{MAP}_e(\epsilon)$ also for the
MAP decoder, from the area theorem of \cite{Ashikhmin:AreaTheorem}
the area below $p^\mathsf{MAP}_e(\epsilon)$ equals the ensemble
rate. By noting that for any $\epsilon$,
$p^\mathsf{MAP}_e(\epsilon)\leq p^\mathsf{IT}_e(\epsilon)$, an upper
bound \cite{Measson08:Maxwell} on
$\epsilon^\mathsf{MAP}_\mathsf{block}$ is given by the value
$\bar{\epsilon}^\mathsf{MAP}_\mathsf{block}$ such that
\begin{equation}
\int_{\bar{\epsilon}^\mathsf{MAP}_\mathsf{block}}^{1}
p^\mathsf{IT}_e(\epsilon)d\epsilon=R_0.\label{eq:areath_MAP}
\end{equation}
This allows us also to get an upper bound on the decoding threshold
for a $d$-regular block CSA scheme, under GA-MAP decoding. Letting
$\alpha=N/M=1/(1-R_0)$, the GA-MAP threshold of CSA can be upper
bounded as
\[
\overline{G}^\mathsf{MAP}_\mathsf{block}=\alpha \overline{\epsilon}^\mathsf{MAP}_\mathsf{block}.
\]

\subsection{Threshold Saturation}\label{sec:results} Table \ref{tab:Thresholds}
illustrates the threshold achievable by conventional CSA schemes
employing a regular distribution at the BNs based on $(d,1)$
repetition codes.  For the spatially-coupled scheme, a super-frame
composed by $M_f=l+d-1$ frames has been considered, with $l=200$.
Moreover, the normalized user population size is $\alpha=100$, i.e.
the number of users is $100$ times larger than the number of slots
per frame. We additionally provide the upper bounds on the threshold
achievable by the conventional CSA scheme under the GA-MAP recovery
process. The derivation of the MAP thresholds serves to illustrate
how, also in this context, the imposition of a convolutional-like
structure to the access scheme allows achieving the threshold
saturation effect as numerically shown in Table
\ref{tab:Thresholds}. The upper bound on the achievable threshold
$G^*$ according to \eqref{eq:AREA_CONDITION_FINAL}, given by the
solution of $G=1-\exp(-G/R)$, is provided too. Accordingly, we
evaluated the normalized efficiency of the proposed scheme as
$$\eta=G^\mathsf{IT}_\mathsf{conv}/G^*\, .$$
As already observed in the LDPC context, larger degrees allow to
approach the bound more tightly.

\begin{table}[t]
\begin{center}
\caption{Thresholds of different access schemes, compared with the
upper bound $G^*$.}\label{tab:Thresholds}
\begin{tabular}{ c c c c c c}
   \hline \hline \\ \vspace{1mm}
  $d$ & $G^\mathsf{IT}_\mathsf{block}$ & $G^\mathsf{IT}_\mathsf{conv}$ & $\overline{G}^\mathsf{MAP}_\mathsf{block}$ & $G^*$ & $\eta$ \\   \hline
  $2$ & $0.5$ & $0.5$ & $0.5$ & $0.7969$ &$0.3726$\\
  $3$ &  $0.8184$   &   $0.9179$  & $0.9179$ & $0.9405$  & $0.9760$\\
  $4$ &  $0.7722$   &   $0.9767$  & $0.9767$ & $0.9802$  & $0.9964$\\
  $5$ &  $0.7017$   &   $0.9924$  & $0.9924$ & $0.9931$  & $0.9993$\\
  $6$ &  $0.6370$   &   $0.9973$  & $0.9973$ & $0.9975$  & $0.9998$\\
  \hline  \hline
\end{tabular}
\end{center}
\end{table}


\section{Conclusion}\label{sec:conclusion}
In this paper we introduced a spatially-coupled RA scheme for the
CCw/oFB which attains capacities close to
$1\,\mathrm{[packet/slot]}$ in the asymptotic (i.e., for large
frames) setting. A bridge between the graphical model describing the
iterative interference cancelation process of the proposed RA over
the random access frame and the erasure recovery process of
low-density parity-check codes over the binary erasure channel has
been set, which allows computing an upper bound on the capacity
achievable by an enhanced (genie-aided) decoder. The saturation of
the SIC IT capacity of the proposed scheme towards the threshold
under genie-aided decoding has been numerically demonstrated.


\end{document}